\documentclass[aps,pre,showpacs,floatfix,preprint]{revtex4}
\usepackage{amssymb}
\usepackage{amsmath}
\usepackage{amsfonts}
\usepackage{dcolumn}
\usepackage{graphicx}
\usepackage[T1]{fontenc}
\usepackage[utf8]{inputenc}
\usepackage{subfigure}
\usepackage{color}
\begin{document}

\title{Phase transitions in two-dimensional model colloids in a one-dimensional external potential}

\author{Florian B\"{u}rzle and Peter Nielaba}

\affiliation{Department of Physics, University of Konstanz, 78457 Konstanz, Germany}

\date{\today}

\begin{abstract}
Two-dimensional melting transitions for model colloids in presence of a one-dimensional external periodic potential are investigated using Monte Carlo simulation and finite size scaling techniques. Here we explore a hard disk system with commensurability ratio $p=\sqrt{3}a_s/(2d)=2$, where $a_s$ is the mean distance between the disks and $d$ the period of the external potential. Three phases, the modulate liquid, the locked smectic and the locked floating solid are observed, in agreement with other experimental and analytical studies. Various statistical quantities like order parameters, their cumulants and response functions, are used to obtain a phase diagram for the transitions between these three phases.
\end{abstract}


\maketitle

Phase transitions in systems with reduced dimensionality have been of interest at least since the days of Landau \cite{Landau1937a} and Peierls \cite{Peierls1934,Peierls1935} on whose ideas the work of Kosterlitz, Thouless, Halperin, Nelson, and Young (KTHNY) \cite{Kosterlitz1973,Kosterlitz1974,Halperin1978,Nelson1979,Young1979} is based to a large extent.

Since the formulation of their successful KTHNY theory, two-dimensional (2D) systems with periodic substates received more and more attention. In their groundbreaking work, Chowdhury, Ackerson, and Clark \cite{Chowdhury1985}, first investigated a 2D colloidal system under the influence of a one-dimensional (1D) periodic potential. This was achieved by interference of two laser beams, yielding a 1D interference pattern. At a strong enough light intensity, crystallization of the colloidal suspension was observed, provided that the periodicity $d$ of the periodic potential was chosen to be commensurate to the mean particle distance $a_s$. For this phenomenon, Chowdhury \emph{et al.}\ coined the name ''laser induced freezing'' (LIF). LIF is, qualitatively, due to the suppression of thermal fluctuations transverse to the 1D periodic potential. Surprisingly, a remelting of the crystal at even higher light intensities was also observed in systems with short-range interactions between the colloids \cite{Wei1998,Bechinger2000,Bechinger2001}. Consequentially, this reentrance scenario was named ''laser induced melting'' (LIM). This was explained as generic, caused by suppression of phonon fluctuations transverse to the potential troughs, leading to a decoupling of neighboring rows \cite{Wei1998}. An analytical justification for this explanation was subsequently given in \cite{Radzihovsky2001}.

It should be emphasized that it was also possible to confirm this scenario by simulation, although some earlier simulations were inconclusive with respect to the reentrance phenomenon (for a more extensive discussion see \cite{Strepp2001,Radzihovsky2001}). In particular, Chakrabarti \emph{et al.}\ \cite{Chakrabarti1995a} recognized LIM even before its experimental discovery. Other Monte Carlo studies also verified the reentrance behavior by considering hard and soft disks, as well as Lennard-Jones-like interaction potentials \cite{Strepp2001,Strepp2002,Strepp2002a,Strepp2003}. Also numerical renormalization group studies on this subject have been successfully performed \cite{Chaudhuri2004,Chaudhuri2004a,Chaudhuri2006}.

It turned out, however, that theoretical attempts based on Landau expansion and related mean field techniques failed to provide an explanation for the complete phenomenology (see \cite{Radzihovsky2001} for a short review). The latter is believed to be due to the incorrect treatment of fluctuations in such theories while it is known that, unlike in three dimensions, fluctuations play a vital role in low dimensional systems and therefore cannot be neglected.

This uncomfortable situation did not change until Radzihovsky, Frey, and Nelson \cite{Frey1999, Radzihovsky2001} published their theory which is based on an elastic model of the triangular lattice. Within this framework, all phase transitions considered so far can be explained by dislocation unbinding. As a further key result, the authors predicted also the existence of new unobserved phases. According to their analysis, the appearance of new phases and hence the complexity of the phase diagram depends, with given crystal orientation, only on the value of an integral quantity $p$ which they called \emph{commensurability ratio}. The latter is a generalization of the concept of commensurability and is formally defined by $p=\sqrt{3}a_s/(2d)$, using the notation convention from \cite{Strepp2001}. For $p=1$ this coincides with the previous meaning of commensurability. In this case, the theory predicts two distinct phases - the modulated liquid (ML), which mimics the geometry of the periodic potential, and the crystalline phase, which Radzihovsky \emph{et al.}\ named ''locked floating solid'' (LFS). This name reflects the dual character of this phase, since the colloids are unpinned along the potential minima, but pinned perpendicular to the minima.
The appearance of those two phases and also the predicted shape of the melting curve are in good agreement with observations from most experiments and simulations discussed so far.

The next difficult case is $p=2$, of which we will present a Monte Carlo analysis in this work. For this case, Radzihovsky \emph{et al.}\ predicted an additional phase, the ''locked smectic'' (LSm). This phase is characterized by a spontaneous symmetry breaking of the discrete translational symmetry present in the modulated liquid, with equal occupancy of each potential minima, while in the LSm only every $p$th minima is equivalently populated. The LSm exhibits, in contrast to the LSF, only short-range correlations between colloidal positions in adjacent troughs and therefore does not resist shear deformations for displacements along the potential minima.

According to \cite{Radzihovsky2001}, both the transition from LFS to LSm and from LSm to ML should exhibit a reentrance behavior. Additionally, our investigation was stimulated by the work of Baumgartl \emph{et al.} \cite{Baumgartl2004} who recently observed the LSm phase at $p=2$ experimentally.

For our analysis, we consider a system of hard disks with diameter $\sigma$. They interact via a pair potential defined by
\begin{equation}
\Phi(r_{ij})=\left\lbrace 
\begin{matrix}
\infty, & r_{ij} \leq \sigma  \\
0, &  r_{ij} > \sigma 
\end{matrix}
\right. 
\end{equation}
where $r_{ij}$ is the distance between particles $i$ and $j$. These particles are confined to a two-dimensional box with dimensions $L_x\times L_y$, with $L_x/L_y=\sqrt{3}/2$. This system is subjected to an external potential
\begin{equation}
 V(x,y) = V_0 \sin\left( 2\frac{2\pi}{d_0}x \right) 
\label{eq:potential}
\end{equation}
which is periodic in the $x$-direction and constant in the $y$-direction (note that in \cite{Radzihovsky2001} the periodicity is chosen to be in the $y$-direction). The constant $d_0$ in eq. (\ref{eq:potential}) is defined as $d_0=a_s\sqrt{3}/2$ to meet the requirements from commensurability. Our system is completely characterized by two quantities, namely by the reduced density $\varrho^*=\varrho\sigma^2$ and the reduced potential strength $V_0^*=V_0/(k_BT)$, where $k_B$ is the Boltzmann constant and $T$ the temperature. For simplification, $\sigma$ was set to unity in our simulation.

\begin{figure}[bt]
\subfigure{\includegraphics[width=4.2cm]{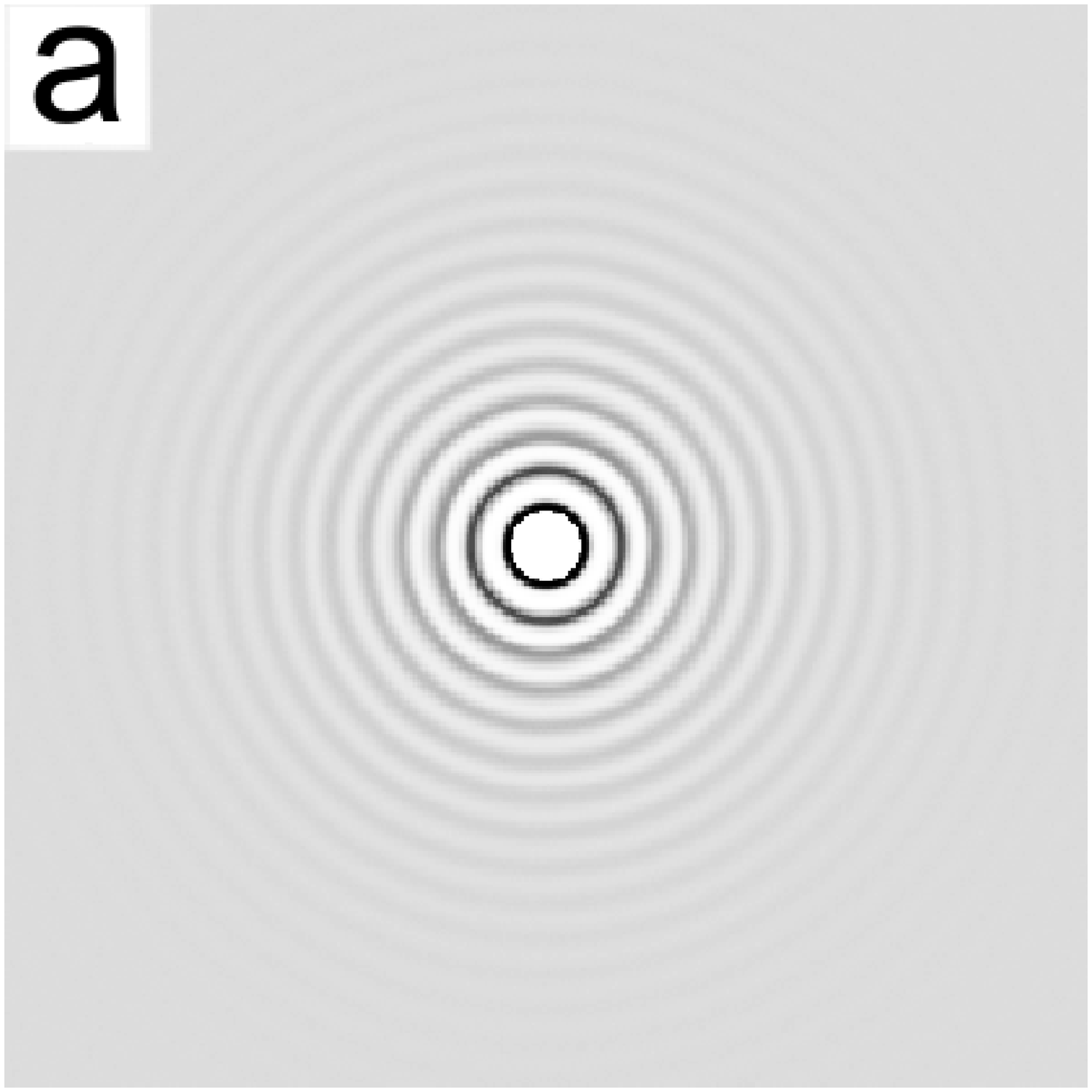}}\hspace*{0.2cm}
\subfigure{\includegraphics[width=4.2cm]{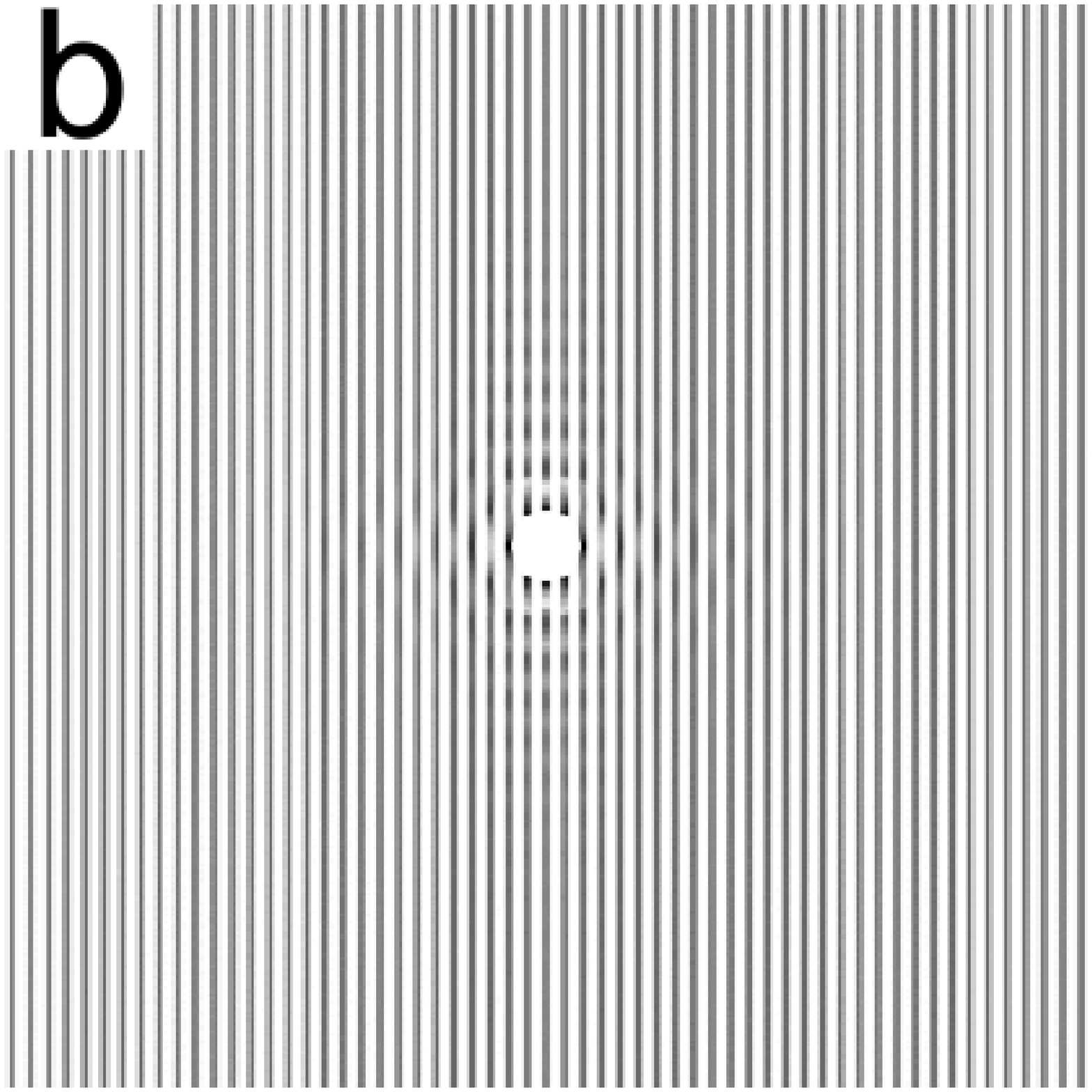}}
\subfigure{\includegraphics[width=4.2cm]{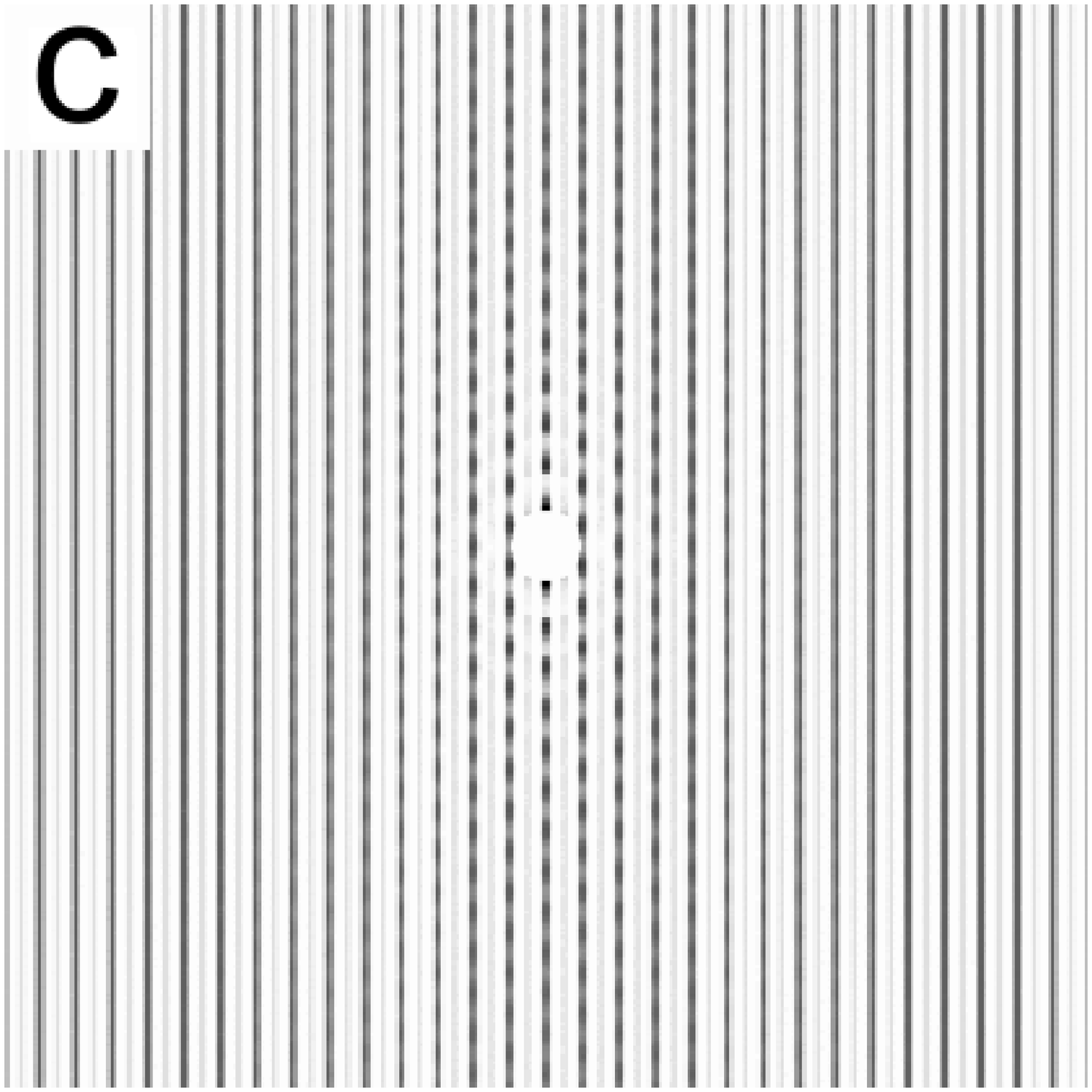}}\hspace*{0.2cm}
\subfigure{\includegraphics[width=4.2cm]{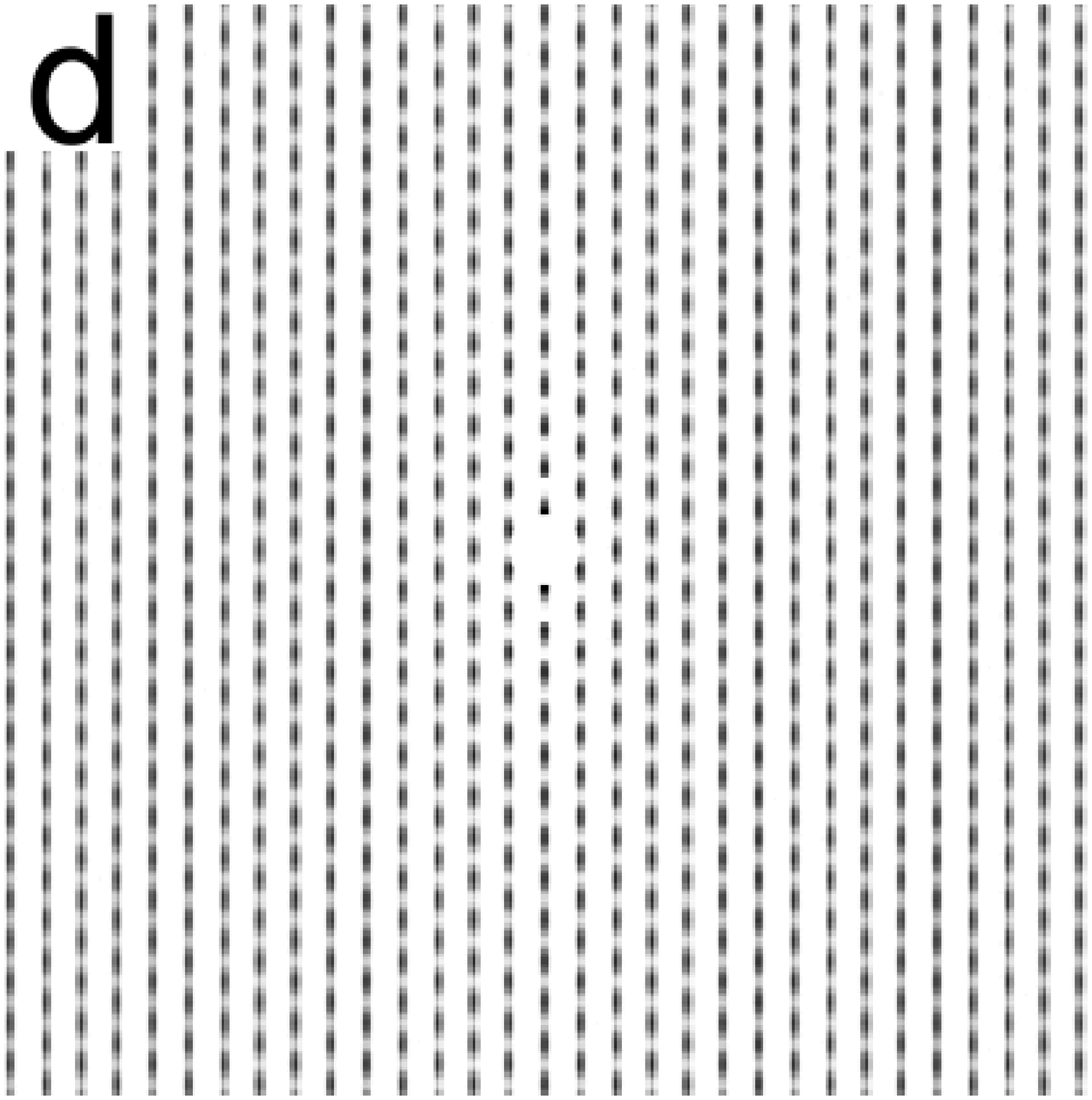}}
\caption{\label{fig:correlation}Two-dimensional pair correlation functions. In (a) an almost unperturbed liquid ($V_0^*=0.01,\rho^*=0.88$) is shown. At the potential strength $V_0^*=5$ the other images show the (b) modulated liquid ($\rho^*=0.83$), (c) locked smectic ($\rho^*=0.86$), and (d) locked floating solid ($\rho^*=0.90$).}
\end{figure}

In this work we use standard Monte Carlo techniques \cite{Landau2005}. In particular, the Metropolis algorithm \cite{Metropolis1953} is applied to the canonical (\emph{NVT}) ensemble. Further we use the technique of block analysis, where we divide our system in several smaller subsystems and calculate the quantities of interest therein. This method allows us to compute many different system sizes within one simulation run. In this paper, we distinguish these subsystems by attaching an index $L$ on the appropriate quantities. The value of $L$ is calculated via $S_x/d$ with $S_x$ being the side length in the $x$-direction of a subbox.

All simulations considered here have been carried out with $N=1024$ particles, in order to compare the results with \cite{Strepp2001}. We executed our calculations by setting up an ordered crystal and reducing $\varrho^*$ at fixed $V_0^*$. For equilibration, especially at high potential strengths, one must ensure that the system under consideration cannot be trapped in local free energy minima. In experiments, this problem can be solved by switching the laser off and on while in simulation this is of course not possible. To overcome this issue, we use nonlocal ''trough moves'', already introduced in \cite{Strepp2001}, in addition to the ordinary Monte Carlo moves. In these trough moves, that are used in every simulation run, it is attempted to place particles in other potential troughs. It was possible to show \cite{Strepp2001}, that such moves are required to reach equilibrium, since at large $V_0^*$ the formation of dislocations is artificially hindered by considering only local moves, because the particles cannot bypass each other.

Our simulations were carried out mainly on contemporary personal computers with $4\times10^7$ to $10\times10^7$ Monte Carlo steps (MCS) from which $2\times10^7$ to $4\times10^7$ were used for relaxation, depending on statistical inefficiency. Typical simulation times ranged from about 50 to 120 CPU hours.

In order to distinguish all appearing phases in our simulation by visual inspection, we computed two-dimensional pair correlation functions. For systems with densities below crystallization and without external potential, the graphs of these functions should exhibit the typical concentric circles of an unperturbed liquid. With an external potential, one expects to see lines along the potential minima characterizing the modulated liquid. However, in the crystalline (LFS) case, the typical structure of the triangular lattice should crop up. In the LSm phase, compared to the modulated liquid, every second line must vanish due to breaking of translational symmetry.

A survey of such 2D correlation functions at different $V_0^*$ and $\varrho^*$ already revealed some interesting properties of the phases. For potential strengths $V_0^*\ll 1$, we found for densities below spontaneous solidification no significant disturbance by the external potential and hence no smectic phase, as one would expect for almost unperturbed liquids [see Fig. \ref{fig:correlation}(a) as typical example]. At intermediate potential strengths with $V_0^*\gtrsim 1$, we found a completely different behavior in the liquid regime. To exemplify this, we consider a potential strength $V_0^*=5$ at various densities $\varrho^*$. At $\varrho^*=0.83$ we recognize a modulated liquid where every potential minima, on the average, is occupied with particles [Fig. \ref{fig:correlation}(b)]. For a slightly higher density at $\varrho^*=0.86$ we find indeed the predicted LSm phase, where only every second lane is occupied [Fig. \ref{fig:correlation}(c)]. Finally, at $\varrho^*=0.90$ the structure of the triangular lattice can be observed, indicating the presence of the crystalline LFS phase [Fig. \ref{fig:correlation}(d)]. We would like to mention, that these graphs are qualitatively in good agreement with Baumgartl \emph{et al.}\ \cite{Baumgartl2004} who obtained similar results for the appropriate phases. For larger potential strengths, it is recognized that the density range in which the LSm phase appears becomes both narrower and shifted to larger densities. We interpret this observation as a strong evidence for a remelting transition in this regime. 

Since the circumstances in the LFS phase are very similar to the appropriate phase in the case $p=1$, we concentrate now on the phase transitions between ML and the new LSm phase. These freezing and remelting processes can qualitatively be described as follows. At low potential strength ($V_0^*\ll1$) the ML is quite disordered, i.\ e.\ the particles are randomly distributed between the potential minima. The increase of the potential strength ($V_0^*\gtrsim1$) then leads to a reduction of fluctuations perpendicular to the potential troughs. Subsequently the particles occupy every second potential minima and form a quasi-long-range orientational order along the troughs. This is in contrast to the LFS, where also a quasi-long-range positional order exists; but it must be noted that still some particles or even particle groups may overcome the potential barrier and occupy free positions in adjacent troughs. This prevents complete decorrelation of neighboring particle rows; but at even higher potential strengths ($V_0^*\gg1$) the fluctuations perpendicular to the troughs are further reduced. Consequentially, particles can occupy adjacent rows without geometric restrictions due to particles from other rows. From this it can be concluded, that the entropy of the system in the ML phase would be higher than in the LSm phase. Since interaction energies are for large potential strengths approximately the same in both phases, the free energy in the ML phase becomes lowered and so a transition from LSm to ML is possible.

For numerical calculation of the phase transitions points, we introduce two different order parameters. Both are Fourier transforms of the particle densities in direction of reciprocal lattice vectors. We consider only the set of the six smallest reciprocal lattice vectors of the 2D triangular lattice. To detect the locked floating solid phase, the vector $\mathbf{G}_1 = 2\pi/d_0 (1/2 \mathbf{e}_x + \sqrt{3}/2 \mathbf{e}_y)$, enclosing an angle of $\pi/3$ with the wave vector $\mathbf{K}=4\pi/d_0\mathbf{e}_x$ of the external potential, is used. With $\mathbf{G}_1$ the definition of the appropriate order parameter $\psi_{\mathbf{G}_1}$ reads
\begin{equation}
\psi_{\mathbf{G}_1} = \left | \sum_{k=1}^N \exp(-i\mathbf{G}_1 \mathbf{r}_k)\right |,
\end{equation}
where $\mathbf{r}_k$ is the position of particle $k$. Note that this definition was first introduced in \cite{Strepp2001} for the treatment of the case $p=1$. For the transition to LSm, we choose the vector $\mathbf{G}_0=2\pi/d_0\mathbf{e}_x$ which is parallel to $\mathbf{K}$ and has half of its magnitude. The corresponding order parameter $\psi_{\mathbf{G}_0}$ is then defined analogous to $\psi_{\mathbf{G}_1}$.

From these order parameters, the phase transition points have been calculated with the cumulant intersection method \cite{Binder1981,Binder1981a}. The fourth order cumulant $U_L$ is defined via
\begin{equation}
U_L\left(\varrho^*,V_0^* \right) = 1 - \frac{\left < \psi_{\mathbf{G}_i}^4 \right >_L }{3\left < \psi_{\mathbf{G}_i}^2 \right >^2_L }
\end{equation}
where the index $i$ has either the value $0$ or $1$, depending on the appropriate order parameter. Since in continuous phase transitions the correlation length $\xi$ diverges, the cumulant $U_L=U_L(La_s/\xi)$ becomes independent of the system size. From this it follows that cumulants for different system sizes intersect at the transition point. Note that there is also an intersection point in first order phase transitions \cite{Vollmayr1993} which makes it unnecessary for us to judge whether the transitions observed are of first order or continuous.

It must be noted that, at high $V_0^*$, the cumulants to both order parameters do not show a true intersection point any more. Instead, there is a unification point, after which all cumulants collapse onto a single curve. This behavior is known to be typical to the anisotropic \emph{XY}-model \cite{Landau1983}, which was also observed in the study by Strepp \emph{et al.}\ \cite{Strepp2001}. In those cases, at this unification point the phase transition was assumed to take place.

Additionally, to supplement our cumulant analysis, we consider the response functions
\begin{equation}
k_BT\chi_{\mathbf{G}_i} = L^2 \left[\left< \psi_{\mathbf{G}_i}^2 \right>_L - \left< \psi_{\mathbf{G}_i} \right>_L^2\right],
\end{equation}
also defined as in \cite{Strepp2001}. These susceptibilities are known to increase with increasing $L$, yielding a maximum in the density range where the phase transition occurs.
Those maxima are shifted to lower densities compared to the cumulant intersection points, due to finite size effects \cite{Strepp2001}. Nevertheless, the form of the melting curve should be in agreement with those obtained with the cumulant intersection method. In our calculations we observe that this is indeed the case. For the sake of clarity, since a comparison with the configurations and 2D pair correlation functions shows that the results obtained by cumulant intersection points are more accurate, the susceptibility maxima have been omitted from the phase diagram.

\begin{figure}
\includegraphics[width=16.4cm]{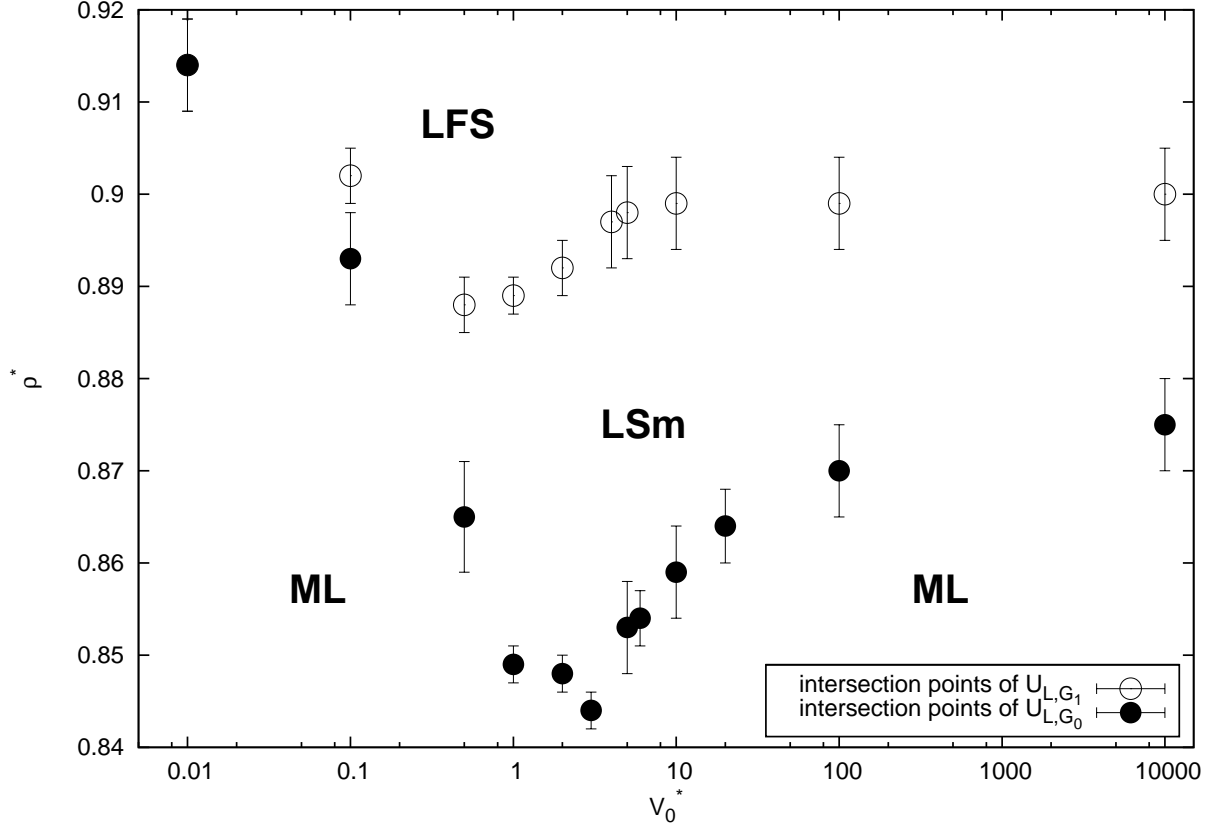}
\caption{\label{fig:pd} Phase diagram in the $\rho^* / V_0^*$ plane. Transitions points have been obtained by considering cumulant intersection points. Open circles: order parameter $\psi_{\mathbf{G}_1}$, and closed circles: $\psi_{\mathbf{G}_0}$.}
\end{figure}

The phase diagram shown in Fig.\ \ref{fig:pd} was obtained by using the data from the cumulant intersection points. As the most important result, we see that the melting curves for both order parameters show a distinct remelting behavior at higher $V_0^*$, as was expected by the theory of Radzihovsky \emph{et al.} By considering the curve composed of cumulant intersection points belonging to $\psi_{\mathbf{G}_1}$ (open circles), we find that the melting curve resembles those obtained by Strepp \cite{Strepp2001} for $p=1$ quite well. The other transition curve from ML to LSm (closed circles) shows that here the global minimum of the curve is slightly shifted to higher potential strengths. Also the minimum is located at considerable lower densities. Finally, it must be emphasized that at $V_0^* \rightarrow 0$, the different melting curves collapse into one single curve, as is expected for physical reasons. 

In conclusion, we have investigated phase transitions in 2D model colloids by Monte Carlo simulation. In particular, melting transitions in the presence of a 1D periodic potential were studied for the commensurability ratio $p=2$. In contrast to the case $p=1$, an additional intermediate phase between the LFS and the ML, the LSm, has been confirmed. This is qualitatively in good agreement with theoretical predictions and experimental results alike. Furthermore, by defining appropriate order parameters, we were able to construct a phase diagram for the three phases observed. This is a substantial progress over the experimental work \cite{Baumgartl2004}, since in that work only isolated points of the whole parameter space were captured.

Finally, we would also like to encourage further experimental work on this exciting subject, especially the construction of a phase diagram for this case experimentally. In addition, explorations of cases with even higher commensurability ratios, where further phases should appear, would be of great interest.

We want to thank our former colleague Wolfram Strepp for his preliminary work on the subject and for providing us with the necessary details. This work was, in part, supported by the Deutsche Forschungsgemeinschaft (SFB TR6/C4). Granting of computer time from HLRS, NIC, and SSC is gratefully acknowledged.

\bibliography{refsnew}

\end{document}